\documentclass[runningheads]{llncs}

\usepackage[T1]{fontenc}
\usepackage[utf8]{inputenc}
\usepackage{graphicx}
\usepackage[frozencache=true,cachedir=.]{minted}
\usepackage{hyperref}
\usepackage{color}
\usepackage{colortbl}
\usepackage{./gospel}
\usepackage{xspace}

\newcommand{\whyml}{\textsf{WhyML}\xspace}

\newcommand{\ocaml}{\textsf{OCaml}\xspace}

\newcommand{\why}{\textsf{Why3}\xspace}

\definecolor{thegray}{rgb}{0.9,0.9,0.9}
\definecolor{colorspec}{rgb}{0,0,0.797}
\definecolor{thered}{rgb}{0.797,0,0}
\definecolor{darkgreen}{rgb}{0.797,0,0}
\definecolor{theblue}{rgb}{0,0,0.797}
\definecolor{darkgray}{rgb}{0.8477,0.8477,0.8477}
\definecolor{ocaml-bg}{rgb}{0.9,0.9,0.9}
\definecolor{thegraygray}{rgb}{0.5,0.5,0.5}

\begin{document}

\definecolor{thegray}{rgb}{0.9,0.9,0.9}
\definecolor{colorspec}{rgb}{0,0,0.797}
\definecolor{thered}{rgb}{0.797,0,0}
\definecolor{darkgreen}{rgb}{0.797,0,0}
\definecolor{theblue}{rgb}{0,0,0.797}
%\definecolor{thegray}{rgb}{0.949,0.949,0.949}
\definecolor{darkgray}{rgb}{0.8477,0.8477,0.8477}
\definecolor{ocaml-bg}{rgb}{0.9,0.9,0.9}
\definecolor{thegraygray}{rgb}{0.5,0.5,0.5}

\newcommand{\cameleer}{\textsf{Cameleer}\xspace}
\newcommand{\GOSPEL}{{\textsf{GOSPEL}}\xspace}
\newcommand{\pre}{precondition\xspace}
\newcommand{\post}{postcondition\xspace}

\title{Auto-active Verification of Graph Algorithms, Written in
  \ocaml\thanks{This work is partly supported by the HORIZON 2020 Cameleer
    project (Marie Skłodowska-Curie grant agreement ID:897873) and NOVA LINCS
    (Ref. UIDB/04516/2020)}}

\author{Daniel Castanho \and Mário Pereira}

%\authorrunning{D. Castanho \and Mário Pereira}

\institute{NOVA LINCS -- NOVA School of Science and Technology, Lisbon, Portugal}

\maketitle

\begin{abstract}
  Functional programming offers the perfect ground for building
  correct-by-construction software. Languages of such paradigm normally feature
  state-of-the-art type systems, good abstraction mechanisms, and well-defined
  execution models. We claim that all of these make software written in a
  functional language excellent targets for formal certification. Yet, somehow
  surprising, techniques such as deductive verification have been seldom applied
  to large-scale programs, written in mainstream functional languages. In this
  paper, we wish to address this situation and present the auto-active proof of
  realistic \ocaml implementations. We choose implementations issued from the
  OCamlgraph library as our target, since this is both a large-scale and
  widely-used piece of \ocaml code. We use \cameleer, a recently proposed tool
  for the deductive verification of \ocaml programs, to conduct the proofs of
  the selected case studies. The vast majority of such proofs are completed
  fully-automatically, using SMT solvers, and when needed we can apply
  lightweight interactive proof inside the \why IDE (\cameleer translates an
  input program into an equivalent \whyml one, the language of the \why
  verification framework). To the best of our knowledge, these are the first
  mechanized, \emph{mostly-automated} proofs of graph algorithms written in
  \ocaml.
\end{abstract}

\section{Introduction}
\label{sec:introduction}

The Verified Software Initiative~\cite{vsi2009} launched an ambitious program of
bringing formal verification techniques to real world software. In other words,
this research manifesto aimed for the stars: to build the foundations of a world
where almost any piece of software could be verified. Over the course of years,
verification tools have emerged from obscure academic artifacts into industrial
standards. Several large companies are, progressively, starting to include
formal methods into their production life
cycle~\cite{ohearn2018,10.1007/978-3-319-96145-3_3}. The so-called \textit{SMT
  revolution} is a key factor in the recent success of formal techniques at an
industrial scale, as automated approaches are more easily integrated within
large code bases. Also, automatic verification is more likely to be adopted by
the regular programmer, when compared with interactive counterparts.

Despite all the advances in deductive verification and proof automation, these
have been seldom applied to the family of functional
languages~\cite{regis-gianas-pottier-08}. In this paper, we wish to remedy the
situation, and at the same time make a contribution to the general effort of
building a corpus of large-scale verified software. We present the formal proof
of realistic, widely-used programs written in \ocaml, a mainstream functional
language. Our tool of choice is \cameleer~\cite{pereiracav21}, a recently
proposed infrastructure for the deductive verification of \ocaml-written code.

We choose to tackle the verification of \ocaml modules issued from the
OCamlgraph library~\cite{DBLP:conf/sfp/ConchonFS07}. The rational for this
choice is twofold: first, this a widely-used library, making its correctness and
safety of utmost importance; second, graph implementations are complex pieces of
software, making it a very interesting challenge for deductive verification. Our
choice might come as a surprise, since OCamlgraph has been extensively used,
hence tested in real-world scenarios. The fact is that recent research has shown
that even massively used libraries are not ironclad, and can hide very subtle
bugs~\cite{brokenTimsort}. OCamlgraph is no exception, as demonstrated by the
issues reported on the project's
repository\footnote{\url{https://github.com/backtracking/ocamlgraph}}.

\paragraph{Contributions.} We present the formal verification of graph data
structures, as well as graph-manipulating algorithms, written in \ocaml. We use
OCamlgraph as our main source of inspiration, and verify the following modules:
\begin{itemize}
\item a persistent data structure representing graphs, with labeled edges;
\item an imperative alternative implementation, with unlabeled edges;
\item a function that checks whether a graph contains a cycle;
\item a breadth-first routine to decide whether there is a path between two
  vertices, in a given graph;
\item an alternative, more efficient and easier to verify, implementation of the
  said path check algorithm.
\end{itemize}
For the mentioned cycle finding function, instead of directly proving the
OCamlgraph implementation, which only returns a Boolean value, we take a
different route: we instrument such function to a \emph{certificate} of the
cycle, \emph{i.e.}, the list of vertices that compose form the cycle. As such,
we then provide a formally verified certificate checker, that ensures the
returned path is indeed a cycle.

Our development differs only slightly from the original OCamlgraph
presentation. Our proofs are modular, leveraging on the \ocaml module system,
preserving a key aspect of the OCamlgraph library, \emph{i.e.}, its clear
separation between algorithms and underlying graph data structures. Since
\cameleer is under active development, this work doubled as a stress test to the
tool and represents the largest \cameleer development to date. Our entire
development (examples, proofs, and proof sessions) is publicly available online
on the project's GitHub
page\footnote{\url{https://github.com/dCastanho/algocameleer}}.

\section{\cameleer in a nutshell}

As mentioned before, \cameleer is a code verification tool, specifically designed
to prove \ocaml code. It is built on top of Why3~\cite{why3-2013}, a set of tools
to perform deductive verification of programs. Why3 comes with its own logic
language, called \textbf{Why}, which is an extension of first-order logic
(e.g. adds recursive definitions, algebric data types, inductive predicates),
and its own programming language, called \textbf{WhyML}. A distinctive feature
provided by this set of tools is that it allows the use of external solvers in
its proofs.

To use \cameleer, we attach \GOSPEL~\cite{gospel} notation comments to our
\ocaml code and feed it to the tool. \GOSPEL is a behavioral specification
language for \ocaml and it is designed to enable modular verification of data
structures and algorithms. These comments represent the specification of our
code and can contain constructs such as preconditions, post-conditions,
variants, invariants. We can also define predicates and functions using \GOSPEL
comments, which come in handy during our proofs. Here is a small example of a
function written in \ocaml, with \GOSPEL comments, which was proved using
\cameleer:

\begin{minted}[linenos]{ocaml}
(*@ open Power *)

let rec fast_exp x n =
  if n = 0 then 1
  else let r = fast_exp x (n / 2) in
    if n mod 2 = 0 then r * r else r * r * x
(*@ r = fast_exp x n
      requires 0 <= n
      variant  n
      ensures  r = x ^ n *)
\end{minted}

The function exemplified above shows a recursive and efficient way to compute a fast exponential, which uses some tricks to calculate the exponential using only $O(log(n))$ multiplications instead of $O(n)$.

Notice the comments started with a \texttt{@} in the above code. These are the \GOSPEL comments mentioned before and define the function's specification, which in this case involves one precondition (the \texttt{requires} clause), one post-condition (the \texttt{ensures} clause), and one variant. We can also use external modules in our proofs (though these modules can only be used in specifications) such as the \texttt{Power} module opened at the start of the proof, which allows the use of the \texttt{\^} symbol.

Here, we tell our user, that if \texttt{n} is equal or greater than 0, then the result will be \texttt{x} to the power of \texttt{n}. Besides indicating what the purpose of the function is, we must assure that it is safe - as in, it does not get stuck in an infinite loop (recursive or otherwise). For this purpose we add the variant, a value that decreases with each call until it reaches a point where the loop ends.

\begin{figure}
\centering
    \includegraphics[scale=0.28]{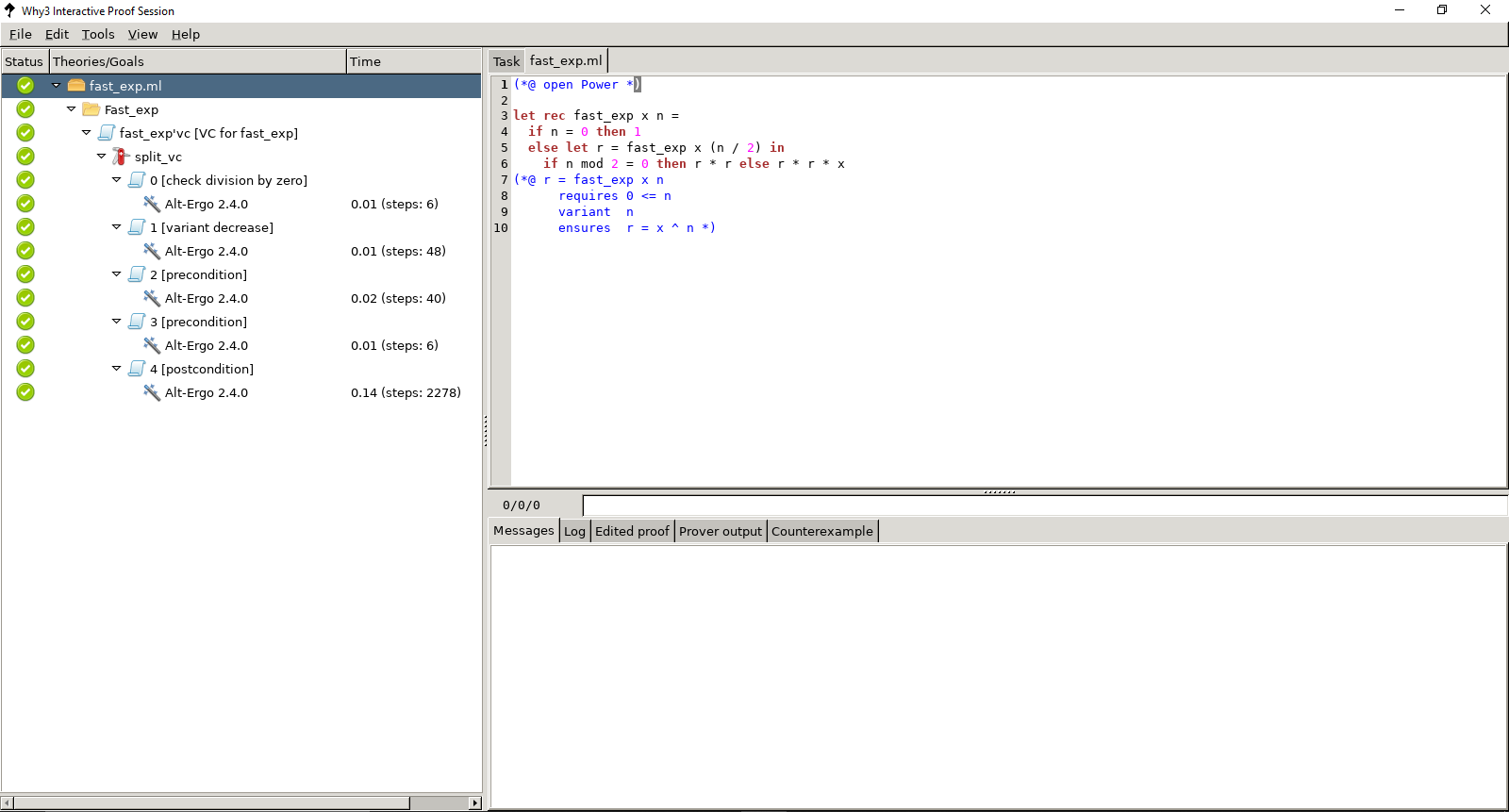}
    \caption{Visualization of the \texttt{fast\_exp.ml} proof in the \textbf{Why3 IDE}}
    \label{fig:fast_exp_label}
\end{figure}

A small thing to keep in mind is that post-conditions can only be assumed to be true if the preconditions are also true when the function is called. For example, if we were to call \texttt{fast\_exp} with a negative \texttt{n}, we could not expect the result to be valid, since our call does not respect the precondition.

Feeding a file to our tool is simple: after installing Cameeler one must simply call the command \texttt{cameleer <filename>.ml}. A graphical interface will pop up, in which all the methods with \GOSPEL specification will be listed and can be selected to be proven. When selected, it is also necessary to choose a theorem prover already installed. In the case of our example we used Alt-Ergo\footnote{\url{https://alt-ergo.ocamlpro.com/}}, which failed to complete the proof right away. Sometimes, it is necessary to split the conditions of a specification using the command \texttt{split\_vc}, which separates the various conditions to try and prove them one by one, assuming the rest are true. After this, Alt-Ergo quickly dispatched the proof of our little example. See \textbf{Figure 1} for a visualization of this example in the \textbf{Why3 IDE}.

\cameleer is still in development at this point, and so it lacks compatibility with some aspects of the core language, such as higher-order functions with side effects or the inclusion of modules in other modules.

\section{OCamlgraph}

As aforementioned, OCamlgraph is a graph library written in \ocaml.  The first notable aspect of this library is that it "does not provide a single data structure for graphs but many of them, enumerating all possible variations (19 altogether)—directed or undirected graphs, persistent or mutable data structures, user-defined labels on vertices or edges, etc.—under a common interface"~\cite{ocamlgraph}. This is possible due to \ocaml's flexible and powerful module system.

This system offers special modules, called functors\footnote{\url{https://ocaml.org/manual/moduleexamples.html\#s\%3Afunctors}} which are important for the second notable feature of this library: the separation of algorithms and data structures, meaning that algorithms are built independently from the underlying data structure. This makes it so, for instance, it does not matter how the vertices and edges are stored, as long as the necessary functions for using the algorithm are supplied in the functor's argument.

OCamlgraph provides a large level of abstraction with all its modules and makes using the library itself incredibly easy, especially since all the graphs end up having very similar signatures. However, analyzing the code behind it proved to be quite difficult because of this abstraction. Not only that, the existence of higher-order functors - functors that instead of returning a module, return another functor - was quite troublesome, not just for us who were inspecting the code, but for \cameleer itself, that was not prepared to handle things of the sort, as mentioned before. Because of this, we were forced to dial down the abstraction, creating a single functor that represented one structure, for each structure analyzed.

\begin{minted}[linenos]{ocaml}
include Graph
include Path

module IntComparable = struct
  type t = int
  let compare a b = if a = b then 0 else if a < b then -1 else 1
  let equal a b = a = b
  let hash a = a
  let default = 0
end

module DP = Persistent.Digraph.
    ConcreteLabeled(IntComparable)(IntComparable)

module CheckPathDP = Path.Check(DP)
let () =
  let g = DP.empty in
  let g = DP.add_vertex g 1 in
  ...
  let g = DP.add_edge g 2 3 in
  let finder = CheckPathDP.create g in
  if CheckPathDP.check_path finder 1 3
      then print_string "Path exists"
      else print_string "Path does not exist"
\end{minted}

In the simple example shown above, we defined a module (\texttt{IntComparable}) that represents an integer, adhering to the \texttt{COMPARABLE} and \texttt{ORDERED\_TYPE\_DFT} signatures, so it can be used as our vertices and edge labels. Afterward, we instantiate our graph's functor using the module we just defined. Since this instantiated module adheres to the signature necessary for \texttt{Path.Check}, we can simply feed it as an argument to the functor. With our modules ready, we can start building our graph. Notice how each call to a function of \texttt{DP} returns a new graph, this is what we call a \emph{persistent} structure in OCamlgraph. There are also \emph{imperative} structures in OCamlgraph, which alter the structure in memory and return \texttt{unit}. If we were to change the type of graph to be used in this example, we would most likely not have to change anything in our code besides the instantiating of the module \texttt{DP} - so long as the type of structure remained the same - because the signatures across graphs are extremely similar.

\section{Formally verified case studies}

Over the course of this work, we proved four different modules. Two of them
represent data structure implementations: the first is an imperative, unlabeled
digraph\footnote{\scriptsize
  \url{https://github.com/dCastanho/algocameleer/blob/main/proofs/imperative_unlabeled_digraph.ml}}
and the second is a persistent, labeled graph%
\footnote{\scriptsize\url{https://github.com/dCastanho/algocameleer/blob/main/proofs/persistent_labeled_graph.ml}}. The
other two consist of algorithms over graphs: the first finds a cycle in a given
graph\footnote{\scriptsize\url{https://github.com/dCastanho/algocameleer/blob/main/proofs/find_cycle.ml}},
and the other checks whether a path between two given vertices exists. We
provide two implementations of this second algorithm: the first is the original
implementation\footnote{\scriptsize\url{https://github.com/dCastanho/algocameleer/blob/main/proofs/path.ml}},
and the second is an altered version, which was easier to prove and is slightly
more
efficient\footnote{\scriptsize\url{https://github.com/dCastanho/algocameleer/blob/main/proofs/altered_path.ml}}. The
two structures were selected to try and cover as many of the available options
for graphs (enumerated in Section 3) as possible, both of which use concrete
(and not abstract) values in their vertices and edges for simplicity's~sake.

In this section, we will present small segments of code regarding each module and explanations about them. For our proofs, we used the following theorem provers: Z3\footnote{\scriptsize\url{https://github.com/Z3Prover/z3}}, Alt-Ergo, CVC4\footnote{\scriptsize\url{https://cvc4.github.io}} and EProver\footnote{\scriptsize\url{https://wwwlehre.dhbw-stuttgart.de/~sschulz/E/E.html}}. Eprover is a slightly different prover than the rest but proved itself to be extremely useful, as some proofs could only be discharged by it.  It is also possible to find some statistics regarding the execution of the proofs in the Appendix.
\subsection{Persistent Labeled Graph}

As explained before, persistent structures are those where each function creates a new version of the original structure and returns such version with the desired change. One of the proved structures is of such kind. It also differs from the other structure in its edges, where the other is unlabeled and directed. This one has generic labels on its edges and they are undirected, which means they go both ways. This structure is incorporated in the following module:

\begin{minted}[linenos]{ocaml}
module PersistentLabeledGraph(Vertex: COMPARABLE)
                              (Edge: ORDERED_TYPE_DFT)
\end{minted}

The vertices and their operations are defined with the \texttt{COMPARABLE} signature. However, the labels of edges also need to be represented, and for such, we use a functor with two arguments like shown above. The second argument is what we will use to represent our edges and their operations, through the \texttt{ORDERED\_TYPE\_DFT} signature, as follows:

 %\newpage

\begin{minted}[linenos]{ocaml}
module type COMPARABLE = sig
  type t
  val[@logic] compare : t -> t -> int
  (*@ axiom pre_order : is_pre_order compare *)
  val hash : t -> int
  val equal : t -> t -> bool
end

module type ORDERED_TYPE_DFT = sig
    type t
    val [@logic] compare : t -> t -> int
    (*@ axiom pre_order_compare: is_pre_order compare*)
    val [@logic] default : t
end
\end{minted}

These signatures both contain a type and a way to compare elements from it, and any module that implements either one must have the represented functions. Notice the \texttt{[@logic]} tag associated to the \texttt{compare} and \texttt{default} functions - this tag means that these functions can be used in specifications. The \texttt{compare} function here is supposed to define an order between the elements of its signature. We guarantee this through the axiom inside \GOSPEL comments, which indicates us that the \texttt{compare} function does represent an order and does not return random numbers (the predicate \texttt{is\_pre\_order} can be found in the standard \GOSPEL library). This assurance is important because, without it, we could not assume anything related to the \texttt{compare} function. The definition and invariants of the graph itself are interesting subjects to present:

\begin{minted}[linenos]{ocaml}
module HM = Map.Make(Vertex)

module VE = struct
  type t = Vertex.t * Edge.t
  let [@logic] compare (x1, y1) (x2, y2) =
    let cv = V.compare x1 x2 in
    if cv != 0 then cv else Edge.compare y1 y2
  (*@ axiom pre_order_compare: is_pre_order compare*)
end

module S = Set.Make(VE)

type t = { self : S.t HM.t }
(*@ invariant forall v1. Set.mem v1 self.HM.dom ->
    forall e. Set.mem e (self.HM.view v1).S.dom ->
    let (v2, l) = e in
      Set.mem v2 self.HM.dom /\
      Set.mem (v1, l) (self.HM.view v2).S.dom *)
\end{minted}

Here, we introduce a module \texttt{VE} which is used for hashing of vertex and edge pairs. The graph itself is represented as an adjacencies list - each vertex has a set associated to it, representing its outgoing edges. Our sets contain pairs \texttt{(Vertex.t, Edge.t)} which store the destination and the label of the edge. This means parallel edges are allowed, as long as they have different labels.

The type takes on the format of a record, however, originally, it was represented as an alias type, \mintinline{ocaml}{type t = S.t HM.t}. Because type invariants can only be applied to record types, we altered it. The invariant guarantees two things: that all the successors of a vertex also belong to the graph and that if there is an edge \texttt{(v2, l)} in \texttt{v1}'s adjacency list, then the edge \texttt{(v1, l)} exists in \texttt{v2}'s adjacency list. This captures the undirectness of the graph.

In both our structures, we could have chosen to keep the graph type an alias type and not have assigned it an invariant. But to make sure these important invariants were maintained, we would have had to add them as pre and post-conditions to every function that manipulates the graph. We chose to add the invariant for two reasons: first, it is safer, as we do not run the risk of forgetting to add these conditions; secondly, it was a good opportunity to demonstrate how invariants are used and see how \cameleer would handle them in this situation.

Most proofs in the structure were automatically discharged by at least one of our provers. However some times we were forced to interactively  help it along in some more complicated assertions. An example of one such proof was in the case of the \texttt{find\_all\_edges} function, presented in \textbf{Figure 2}. No prover was able to conclude that every member of the list  \mintinline{ocaml}{S.elements (HM.find v1 g.self)} also belonged to the graph. This list represents the successors of the vertex \texttt{v1} and so, according to our invariant, they must belong to the graph. By instantiating the invariant with the graph \texttt{g} and vertex \texttt{v1} as well as the destruction of some logical implications (splitting the premise and the condition) \cameleer managed to reach this same conclusion.

\begin{figure}
\centering
    \includegraphics[scale=0.28]{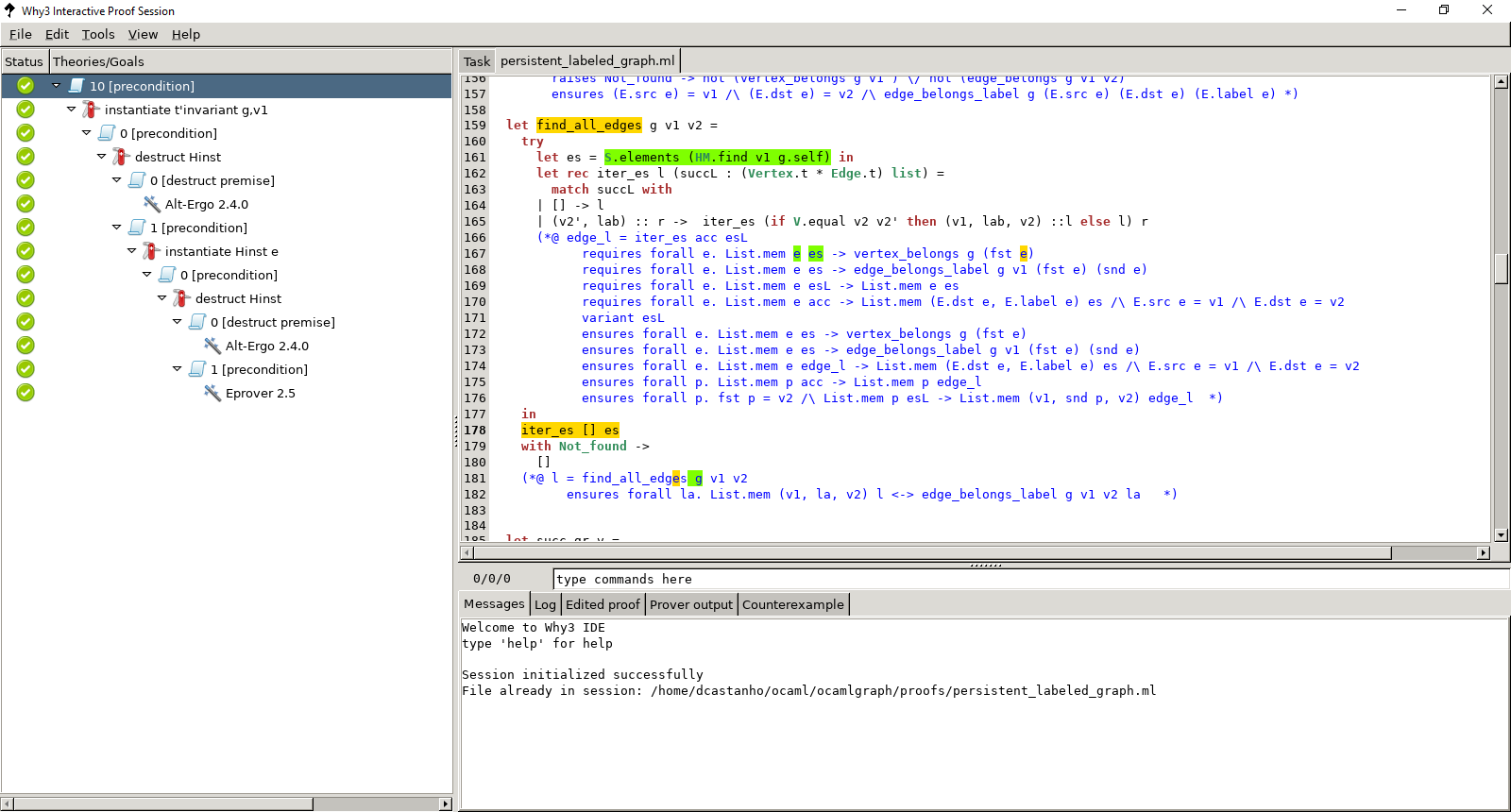}
    \caption{Visualization of the interactive proof for in the \textbf{Why3 IDE}}
    \label{fig:persistent_label}
\end{figure}

Despite many of the proofs being automatic, we had to add some lemmas that look very simple, because some times the provers were having hard time reaching those simple conclusions amidst the proof. When generalized, as a lemma, automatic theorem provers were quick to prove them correct. A small example is the \texttt{eq\_seq\_list} shown ahead, which simply states that all elements of a list are also represented on the sequence generated from it.

\lstset{numbers=left,numberstyle=\tiny}
\begin{gospel}
(*@ lemma eq_seq_list : forall l :  (Vertex.t * S.t) list, e.
                        List.mem e l -> Seq.mem e (of_list l) *)
\end{gospel}

\subsection{Cycle finding}

One of OCamlgraph's cycle detection routines is a function called
\texttt{has\_cycle}, which returns a Boolean, whether or not the graph has a
cycle. Interestingly, one of the tasks on OCamlgraph's to-do list, since around
2015, is updating this function to return the cycle that was found, instead of
simply a Boolean. This led us to open a pull request on the OCamlgraph GitHub
repository, where we propose to add a new function called \texttt{find\_cycle},
inspired by \texttt{has\_cycle}, which returned the cycle found in the form of a
list\footnote{https://github.com/backtracking/ocamlgraph/pull/116}.

Given the complexity of the said \texttt{find\_cycle} function, we decided to
use it for a different kind of proof, one we call \emph{proof through a
  certificate}. This kind of proof is useful when the function or program we are
trying to prove is complex, but given the output of it, we can check whether
that output is correct or not. And so, instead of proving the function itself,
we construct a new support function, a \emph{certificate generator} or
\emph{witness validator} so to say, which receives an output of the first
function and tests whether it is correct, creating a certificate for it. Then we
prove this support function is correct. However, we still prove that the
\texttt{find\_cycle} function is safe, \emph{i.e.}, it always terminates and
does produce any exceptional behavior (\emph{e.g.}, unhandled raised
exceptions).

This is exactly what we did. After adding the new \texttt{find\_cycle} function,
we created a function that given a list and a graph, checks if that list
represents a cycle in the graph. Finally, we proved this function did exactly
that. An interesting feature of this kind of certificate proof, is that our
support function is largely independent of the implementation of the cycle
finder used, so long as the returned list is in the format we expect. As
mentioned previously, all our algorithms are independent of a specific graph
implementation, relying on receiving it as a module that adheres to a certain
signature, as follows:

\begin{minted}[linenos]{ocaml}
module type T = sig
  val is_directed : bool
  module V : COMPARABLE
  type gt
  (*@ model dom: V.t fset
      model suc: V.t -> V.t fset
      invariant forall v1, v2. Set.mem v1 dom /\
        Set.mem v2 (suc v1) -> Set.mem v2 dom*)
   val [@logic] succ : gt -> V.t -> V.t list
   (*@ l = succ g v
        requires Set.mem v g.dom
        ensures forall v'. List.mem v' l -> Set.mem v' g.dom
        ensures forall v'. List.mem v' l <-> Set.mem v' (g.suc v) *)
  val all : gt -> V.t list
  (*@ l = all g
        ensures forall v. List.mem v l <-> Set.mem v g.dom *)
end
\end{minted}

Here, that signature requires the following: a module \texttt{V} that adheres to \texttt{COMPARABLE} and is used to represent the vertices of the graph; a type \texttt{gt}, which represents the type of our graph; a function \texttt{succ}, which given a graph and a vertex, returns its list of successors, and a function \texttt{all}, which given a graph returns a list of all its vertices. Originally, both functions provided higher-order iterators to iterate over these lists, but due to \cameleer's current incompatibility with these, we changed them to return the lists so we could iterate over them manually.

For our specification to make sense of what the \texttt{gt} type represents, we assign a model to it. In this case, our model consists of a finite set (\texttt{fset}) \texttt{domain}, which stores the vertices that belong to the graph and a \texttt{succ} function, that given a vertex, returns its successors set. This model is used to provide a specification of the functions of the signature.

\begin{minted}[linenos]{ocaml}
let is_path v1 l v2 g =
    let rec is_succ v = function
        | [] -> false
        | v' :: vs -> G.V.equal v' v || is_succ v vs
    (*@ b = is_succ v l
        variant l
        ensures b <-> List.mem v l *)
    in let rec iter_path = function
        | [] -> assert false
        | v' :: v'' :: vs -> is_succ (v'') (G.succ g v') &&
                iter_path (v'' :: vs)
        | v' :: [] -> G.V.equal v' v2
    (*@ b = iter_path l
        requires l <> []
        requires forall v. List.mem v l -> Set.mem v g.G.dom
        variant l
        ensures let s = of_list l in
          b <-> (forall i. 0 <= i < List.length l - 1 ->
           edge s[i] s[i+1] g) /\ s[List.length l - 1] = v2 *)
    in match l with
        | [] -> G.V.equal v1 v2
        | x :: _ -> is_succ x (G.succ g v1) && iter_path l
    (*@ b = is_path v1 l v2 g
        requires Set.mem v1 g.G.dom
        requires Set.mem v2 g.G.dom
        requires forall v. List.mem v l -> Set.mem v g.G.dom
        ensures b <-> is_path v1 (of_list l) v2 g *)
\end{minted}

Cycles, by definition, are paths from a vertex to itself. As such, the simplest way to define a cycle is to start by defining a path. The function demonstrated above receives a list and two vertices and tests whether the list represents a path from \texttt{v1} to \texttt{v2}. The list itself, does not contain \texttt{v1} but its last element must be \texttt{v2}. We make the verification by first checking whether the head of the list is a successor of \texttt{v1}. Then, we take each vertex from the list and check if the one that follows it (in the list) is one of its successors (this is done by the \texttt{iter\_path} function). If the list is empty, then it means there are no other vertices involved, so it returns true if \texttt{v1}  and \texttt{v2} are the same vertex.

Our function has several preconditions, which can be summed up by saying that all vertices involved must belong to the graph, otherwise there is no doubt that it is not a path. The function assures us that the returned Boolean is equivalent to the one returned by the predicate of the same name, defined as follows:

\lstset{numbers=left,numberstyle=\tiny}
\begin{gospel}
(*@ predicate edge (v1 : G.V.t) (v2 : G.V.t) (g : G.gt) =
        Set.mem v2 (g.G.suc v1) *)

(*@ predicate is_path (v1 : G.V.t) (l : G.V.t seq)
    (v2 : G.V.t) (g : G.gt) =
            let len = Seq.length l in
            if len = 0 then v1 = v2 else
              edge v1 l[0] g && l[len - 1] = v2 &&
              Set.mem v1 g.G.dom && forall i : int.
               0 <= i < len - 1 -> edge l[i] l[i+1] g *)
\end{gospel}

It is easy to see a similarity between the two functions, however, the predicate can use the logical quantifier \texttt{forall} as well as operate directly over the model defined in the signature. This function was proved fully automatically by \cameleer, though it is still not the certificate creator we want. The certificate generator was defined as follows:
\begin{minted}[linenos]{ocaml}
let is_cycle l g =
    let rec get_last = function
    | x :: [] -> x
    | x :: xs -> get_last xs
    | [] -> assert false
    (*@ x = get_last l
          variant l
          requires l <> []
          ensures List.mem x l
          ensures x = (of_list l)[List.length l - 1] *)
    in not (is_empty l) &&
    let v = get_last l in is_path_func v l v g
(*@ b = is_cycle_func l g
      requires forall v. List.mem v l -> Set.mem v g.G.dom
      ensures b <-> is_cycle (of_list l) g *)
\end{minted}
Now we have to decide how a cycle would be stored in a single list. We ended up by choosing the simplest possible representation: given a list, it is a path from the last vertex to itself (though technically, there is a path from any of its vertices to itself). Given this way of storing cycles, an empty list represents the non-existence of a cycle (as in, empty lists cannot be cycles). The function above does exactly this verification, once again requiring all its members to be part of the graph.

\lstset{numbers=left,numberstyle=\tiny}
\begin{gospel}
 (*@ predicate is_cycle (l : G.V.t seq) (g : G.gt) =
      let len = Seq.length l in
      len <> 0 /\ is_path l[len - 1] l l[len - 1] g*)
\end{gospel}

 The predicate used to represent this is once again very similar to the actual function, simply calling \texttt{is\_path} with the correct values. All these functions were proved to adhere to their specifications by \cameleer, meaning that given any function which \emph{supposedly} returns a cycle, we can check whether their result is correct or not.

 Since this function is something the OCamlgraph developers have had in their to-do list for a while, we made a pull request to add our implementation of it to the library. We also raised an interesting point regarding cycle finding - in the case where a cycle is not found, returning a \emph{correct} topological sort of the graph can also serve as proof that there are no cycles.

\subsection{Path checking}

For simple path checking between two vertices, OCamlgraph has a type and a function, where the type serves as a cache for previously searched values and the function uses breadth-first search to find the path between two given vertices, \texttt{v1} and \texttt{v2}. In this section we will be explaining the original version of the code\footnote{\url{https://github.com/backtracking/ocamlgraph/blob/master/src/path.ml}}.

The implementation is simple but requires two support structures: a queue (called \texttt{q}) for vertices to check and a Hash table (called \texttt{visited}) for checked vertices. We start the algorithm by adding \texttt{v1} to \texttt{q}. At each execution of the algorithm, we check if the queue is empty, if it is, we stop, returning false - as there are no more reachable vertices from \texttt{v1} - if it is not, we pop the first value, \texttt{v}. If the popped value is equal to \texttt{v2}, we have found a path and return true, if not, we add \texttt{v} to \texttt{visited} and iterate over its successors, adding them all to the queue.

There were several conditions to be treated as invariants over the execution of the algorithm, but due to the use of recursive functions instead of loops, these were adapted as pre and post-conditions. The definition of a path is defined just like in the previous subsection regarding cycles. However this time we will provide a more general definition, which instead of being asked if a given sequence is a path, simply checks if there exists such a path.

\lstset{numbers=left,numberstyle=\tiny}
\begin{gospel}
 (*@ predicate has_path (v1 : G.V.t) (v2 : G.V.t) (g : G.gt) =
            exists p : G.V.t seq. is_path v1 p v2 g *)
\end{gospel}

Proving the path was found when the function returns true is easy, thanks to the invariant that states that the vertices in \texttt{q} are all connected to \texttt{v1}, which means if \texttt{v2} is one of them (when popped), then there is a path between the~two.

\lstset{numbers=left,numberstyle=\tiny}
\begin{gospel}
 forall v'. Seq.mem v' q.Queue.view -> has_path v1 v' pc.graph
\end{gospel}

Proving that when the function returns true, then there is a path, refers to the \emph{correctness} of it, while returning false (instead of unsure, for example) when there is no path, refers to its \emph{completeness}. This last one proved to be much more difficult than the first.

In our algorithm, if \texttt{q} is empty when we start a loop, it means there are no more reachable vertices, and so, there is no path between \texttt{v1} and \texttt{v2}. From a programming point of view, this train of thought is natural but logically proving it is a different story. The following invariant is how we prove that if the queue is empty, then there is no path:

\lstset{numbers=left,numberstyle=\tiny}
\begin{gospel}
has_path v1 v2 pc.graph ->
    exists w. Seq.mem w q.Queue.view /\ has_path w v2 pc.graph /\
      not (Set.mem w visited.HV.dom)
\end{gospel}

Because if there is a path, then there must be an intermediate \emph{unvisited} vertex in the queue and if there are no more vertices in the queue, this can't happen. With such a property in hand, we were able to prove the completeness of the function.

However, proving this specific condition was not done automatically and before we did so we needed an extra invariant.

\lstset{numbers=left,numberstyle=\tiny}
\begin{gospel}
 forall v. Set.mem v visited.HV.dom ->
    forall s. edge v s pc.graph ->
        Seq.mem s q.Queue.view \/ Set.mem s visited.HV.dom
\end{gospel}

This means that if a vertex is visited, then all its children have either been visited or are in the queue. \cameleer easily proved this condition and we proceeded to the proof of the previous one, doing so through a case by case analysis:
\begin{enumerate}
    \item If \texttt{v2} is in the queue, then it is easily proven, as it is the vertex we are looking for. The condition \texttt{has\_path v2 v2} trivially holds.
    \item If \texttt{v2} is not in the queue, then we know \texttt{v1} is in \texttt{visited}, because \texttt{v1} is the first vertex to be popped. So we know that the there is a path from inside the \texttt{visited} set (\texttt{v1}) to outside of it (\texttt{v2}) . This path takes an edge \texttt{v}$\rightarrow$\texttt{w}, where \texttt{v} is in  \texttt{visited} and  \texttt{w} is in the queue, but not in visited, and is exactly the vertex we're looking for.
\end{enumerate}

To prove this last case we required the aid of the following ghost function:

\begin{minted}{ocaml}
let [@ghost] [@logic] rec intermediate_value p
    (u : G.V.t) (v : G.V.t) ( s : G.V.t list) ( g : G.gt)  =
      match s with
      | [] -> assert false
      | x :: xs -> if not (p x) then (u, x, [], xs) else
       let (u', v', s1, s2) = intermediate_value_func p x v xs g
       in (u', v', x::s1, s2)
(*@ (u', v', s1, s2 ) = intermediate_value p u v s g
    requires p u /\ not (p v) /\ is_path u (of_list s) v g
    variant s
    ensures p u' /\ not (p v') /\ is_path u s1 u' g
    ensures is_path v' s2 v g /\ edge u' v' g *)
\end{minted}

Given a certain property \texttt{p}, which takes the form of a Boolean function, two vertices, a list representing their path, and the graph where they belong to, it will find the edge in the path which starts in a vertex that respects \texttt{p} and ends in a vertex which does not.

If we instantiate this function with our values \texttt{v1}, \texttt{v2}, our graph, the path used by \texttt{has\_path} and \mintinline{ocaml}{p = fun x -> Set.mem x visited.HM.dom}, it will search for the vertex \texttt{w} referred in the second case. We know this vertex will be in the queue, because we if \texttt{v} (from the edge \texttt{v}$\rightarrow$\texttt{w}) is visited, then all its children are either visited or in the queue. If \texttt{w} is not visited, then it \emph{must} be in the queue.

Because our path definition uses sequences and not OCaml lists , to be able to use this function in our proof correctly (it uses lists and not sequences because sequences are not an \ocaml construct but a \GOSPEL one), we had to create the following auxiliary lemma.

\lstset{numbers=left,numberstyle=\tiny}
\begin{gospel}
  (*@ lemma intermediate_value :
  forall p : (G.V.t -> bool), u, v : G.V.t, s : G.V.t seq, g : G.gt.
    p u -> not p v -> is_path u s v g ->
      exists u' v' s1 s2. p u' /\ not p v' /\
         is_path u s1 u' g /\ is_path v' s2 v g /\ edge u' v' g *)
\end{gospel}

This lemma was what ultimately helped us conclude the proof of completeness and the function explained before was used to prove this lemma. See \textbf{Figure 3} for a visualization of the proof of the condition in the \textbf{Why3 IDE}.

\paragraph{An alternative method.} The way this function is written makes it so the queue and the \texttt{visited} table have overlapping vertices, which forces us to refer to the unvisited but reached vertices by the difference of these two. Not only that, it makes it so the queue might have repeated values and can possibly grow to have more vertices than there are in the graph. We propose an alternative implementation that avoids both of these issues: we made \texttt{visited} a ghost structure and added in a new Hash table by the name of \texttt{marked}; then, instead of adding successors to the queue with no verification, we first check whether that vertex has been marked or not. If it has, we simply move on to the next successor; if it hasn't, we mark it and add it to the queue; on pop vertices are still added to visited. This means that \texttt{visited} and \texttt{q} are disjoint and that all the values in \texttt{q} are distinct. This implementation was also fully proved in \cameleer, with a simpler completeness proof, given the separation of the values in \texttt{visited} and \texttt{q}

\begin{figure}[t]
\centering
    \includegraphics[scale=0.28]{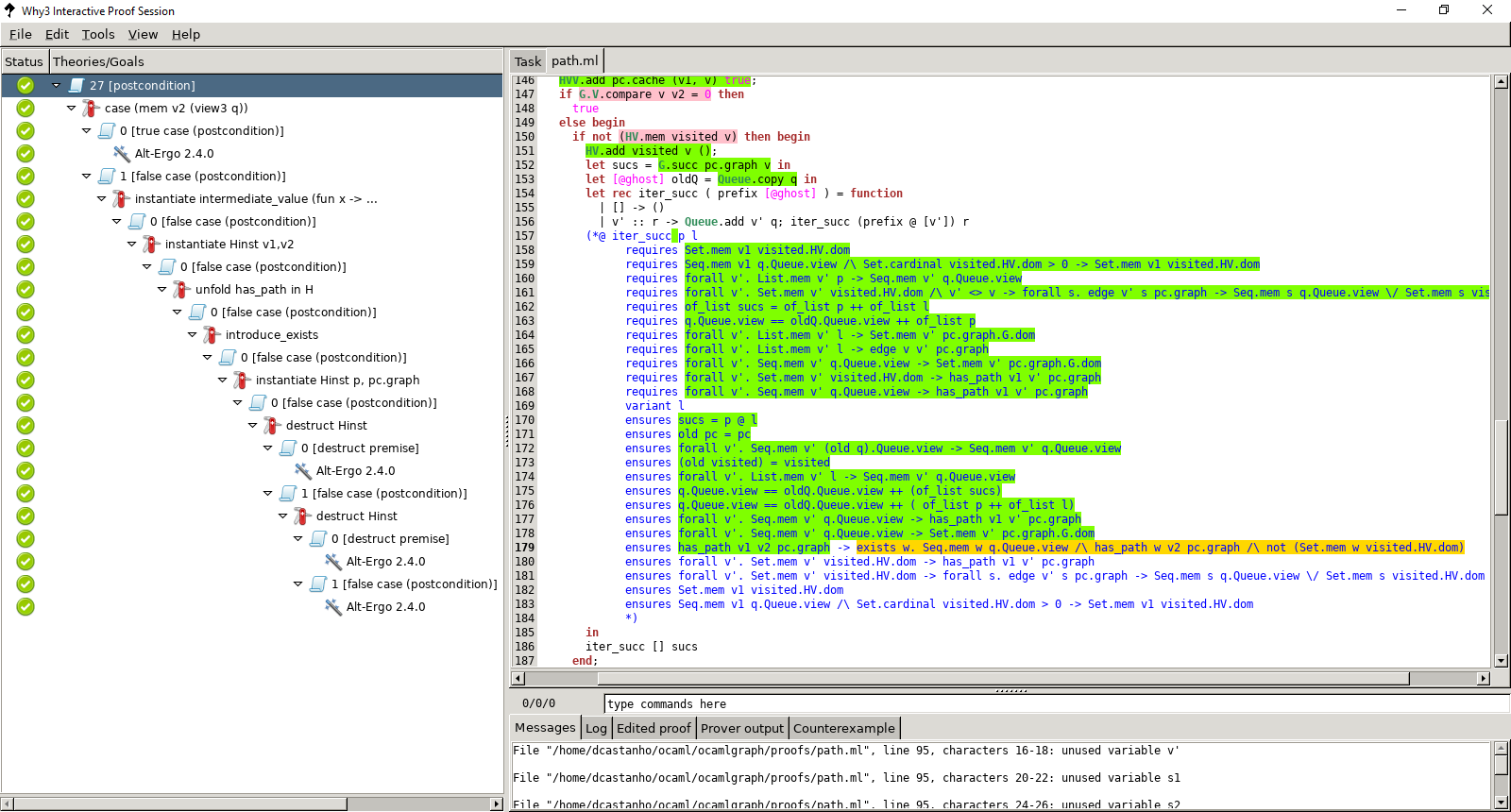}
    \caption{Visualization of the proof of completeness in the \textbf{Why3 IDE}}
    \label{fig:path_label}
\end{figure}

\section{Related work}

Given the importance of graphs, it makes sense that formal verification of algorithms over graphs is not new. For example, the Why3 gallery of verified programs contains a small section specifically for graph algorithms\footnote{\url{http://toccata.lri.fr/gallery/graph.en.html}}, including a breadth-first search such as the one we proved. There are differences, however, such as the fact that their proof was made in WhyML, Why3's programming language, and ours was made in \ocaml. Not to mention that our proof was taken from already written code, instead of a proof-oriented implementation.

Another example of verified graph algorithms is~\cite{graphAlgorithmsC} in which the authors prove some classic algorithms written in C. The implementations proved in this particular work were taken from textbooks and included the verification of heaps and other support structures, along with the rebuttal of some common notions like how \emph{Prim's Algorithm} requires a connected graph.

In~\cite{tarjanAlgorithm} we can find proofs for Tarjan’s strongly connected components algorithm, proved with three different provers: Isabelle/HOL, Coq and Why3.

Alkassar et al.~\cite{certificates} adjust graph algorithms to produce witnesses that can be then used by verified validators to check whether the result is correct, similarly to how we create a certificate and validators for our cycle finding.

Several other works about formal and informal proof of graph algorithms and/or programs exist~\cite{hoareStyle,simplifiedCorrectnessSCC,DFSFramework,pointerPrograms,concurrentPrograms,concurrentGraph,sharingDataStructures,semiAutoSCC}. Some focus on the correct treatment of pointers or concurrent structures while other are part of larger~libraries.

As far as we are aware, there is no \ocaml graph library currently verified and we found no works on the correctness of graph \emph{structures}, only their~algorithms.

\section{Conclusion and Future work}

We presented our verification effort of data structures and algorithms, taking
inspiration from the OCamlgraph library. These proofs were entirely conducted
using the \cameleer tool, with a high-degree of proof automation. We believe our
formal development is an important milestone in the effort of bringing
(automated) verification techniques into real-world software written in \ocaml.

This work led us to extend \cameleer with new features, namely the treatment for
functors instantiation. However, in the grand scheme of things, the subset of
the library we proved is rather small. We intend to keep evolving \cameleer to
handle a larger set of \ocaml programs. As such, our plan for the future
involves expanding on both of these topics. Proving more algorithms, like the
classic \emph{Prim's algorithm}, and structures, like graphs that use abstract
values instead of concrete ones, will increase the verified subset of the
OCamlgraph library. By adding the ability to handle higher-order iterators to
\cameleer, we will fix one of its major flaws and biggest source of
incompatibility with real, already written code. This last one is complicated,
but not impossible, since these iterators are closely related to classic
\texttt{for each} cycles. Following such a route, would allow us to modularly
reason about iteration, as presented by Filliâtre and
Pereira~\cite{pereira16nfm}. From a practical point of view, this means
extending \cameleer with the ability to identify higher-order iterators and
convert them to a piece of WhyML code that uses its enhanced \texttt{for} loop.

\paragraph{Acknowledgments.}We thank António Ravara for his remarks and fruiftul
discussions on draft versions of this paper.

\bibliography{bibliography.bib,local}
\bibliographystyle{splncs04.bst}

\newpage
\appendix

\section{Statistics}

\subsection{Computer and Environment specs}

\begin{itemize}
    \item 16 GB of RAM
    \item AMD Ryzen 5 1600 Six-Core Processor 3.20 GHz
    \item Executed inside WSL 2.0 in Windows 10 64x, with an Ubuntu 20.04 LTS distribution
    \item Why3 setup to have up to 5 processes running at once
\end{itemize}

\subsection{imperative\_unlabeled\_digraph.ml}
\paragraph{Number of root goals.} \textbf{Total:} 22 \textbf{Proved:} 22

\paragraph{Number of sub goals.} \textbf{Total:} 79 \textbf{Proved:} 79

\paragraph{Replay time.} 7 executions timed, removed maximum and minimum. Statistics bellow are from the leftover 5 tests.

\begin{table}
\begin{tabular}{|l|l|l|}
\hline ~\textbf{Minimum}~   & ~\textbf{Maximum}~ & ~\textbf{Average}~~ \\
\hline ~0m15.768s & ~0m16.276s & ~0m16.001s \\ \hline
\end{tabular}
\end{table}

\paragraph{Statistics per prover.} ~

\begin{table}
\begin{tabular}{|l|l|l|l|l|}
\hline
\textbf{Prover}   & \textbf{Number of Proofs} & ~\textbf{Minimum (s)} & \textbf{Maximum (s)} & \textbf{Average (s)}  \\
\hline
Eprover 2.5 &  27 & 0.03 & 2.44 & 0.24 \\
\hline
Alt-Ergo 2.4.0 ~~& 75 & 0.00 & 4.67 & 0.15 \\
\hline
Z3 4.8.7 & 27 & 0.01 & 4.47 & 0.92 \\
\hline
CVC4 1.6  & 40 & 0.06 & 0.25 & 0.12 \\
\hline
\end{tabular}
\end{table}

\newpage
\subsection{persistent\_labeled\_graph.ml}

\paragraph{Number of root goals.} \textbf{Total:} 28 \textbf{Proved:} 28

\paragraph{Number of sub goals.} \textbf{Total:} 219 \textbf{Proved:} 219

\paragraph{Replay time.} 7 executions timed, removed maximum and minimum. Statistics bellow are from the leftover 5 tests.

\begin{table}
\begin{tabular}{|l|l|l|}
\hline ~\textbf{Minimum}~   & ~\textbf{Maximum}~ & ~\textbf{Average}~~ \\
\hline ~0m24.661s & ~0m24.898s & ~0m24.769s \\ \hline
\end{tabular}
\end{table}

\paragraph{Statistics per prover.} ~

\begin{table}
\begin{tabular}{|l|l|l|l|l|}
\hline
\textbf{Prover}   & \textbf{Number of Proofs} & ~\textbf{Minimum (s)} & \textbf{Maximum (s)} & \textbf{Average (s)}  \\
\hline
Eprover 2.5 &  67 &  0.02 & 6.46 & 0.60  \\
\hline
Alt-Ergo 2.4.0 ~~& 179 &  0.01  & 4.71 &  0.13 \\
\hline
Z3 4.8.7 & 55 &  0.01 & 0.22 & 0.07 \\
\hline
Z3 4.8.7 (noBV) &  30 & 0.01 & 0.13 &  0.05 \\
\hline
CVC4 1.6  &  101  & 0.07 &  4.33 &  0.29  \\
\hline
\end{tabular}
\end{table}

\newpage
\subsection{find\_cycle.ml}

\paragraph{Number of root goals.} \textbf{Total:} 9  \textbf{Proved:} 9

\paragraph{Number of sub goals.} \textbf{Total:} 78 \textbf{Proved:} 78

\paragraph{Replay time.} 7 executions timed, removed maximum and minimum. Statistics bellow are from the leftover 5 tests.

\begin{table}
\begin{tabular}{|l|l|l|}
\hline ~\textbf{Minimum}~   & ~\textbf{Maximum}~ & ~\textbf{Average}~~ \\
\hline ~0m18.572s & ~0m19.298s & ~0m18.809s \\ \hline
\end{tabular}
\end{table}

\paragraph{Statistics per prover.} ~

\begin{table}
\begin{tabular}{|l|l|l|l|l|}
\hline
\textbf{Prover}   & \textbf{Number of Proofs} & ~\textbf{Minimum (s)} & \textbf{Maximum (s)} & \textbf{Average (s)}  \\
\hline
Eprover 2.5 & 2 &  0.13 &  0.15 &  0.14 \\
\hline
Alt-Ergo 2.4.0 ~~&   66  & 0.01 &  1.73 &  0.07  \\
\hline
CVC4 1.6  & 6  & 0.11 & 23.71 &  6.86  \\
\hline
\end{tabular}
\end{table}

\newpage
\subsection{path.ml}

\paragraph{Number of root goals.} \textbf{Total:} 14 \textbf{Proved:} 14

\paragraph{Number of sub goals.} \textbf{Total:} 223 \textbf{Proved:} 223

\paragraph{Replay time.} 7 executions timed, removed maximum and minimum. Statistics bellow are from the leftover 5 tests.

\begin{table}
\begin{tabular}{|l|l|l|}
\hline ~\textbf{Minimum}~   & ~\textbf{Maximum}~ & ~\textbf{Average}~~ \\
\hline ~0m22.797s & ~0m22.933s & ~0m22.860s \\ \hline
\end{tabular}
\end{table}

\paragraph{Statistics per prover.} ~

\begin{table}
\begin{tabular}{|l|l|l|l|l|}
\hline
\textbf{Prover}   & \textbf{Number of Proofs} & ~\textbf{Minimum (s)} & \textbf{Maximum (s)} & \textbf{Average (s)}  \\
\hline
Eprover 2.5 &   70 &  0.02 &  3.22 &  0.24 \\
\hline
Alt-Ergo 2.4.0 ~~& 165 &  0.01  & 3.85 &  0.19 \\
\hline
Z3 4.8.7 & 54 &  0.01  & 0.13 &  0.05  \\
\hline
CVC4 1.6  &75 &  0.04 &  5.33 &  0.29  \\
\hline
\end{tabular}
\end{table}

\newpage
\subsection{altered\_path.ml}

\paragraph{Number of root goals.} \textbf{Total:} 15 \textbf{Proved:} 15

\paragraph{Number of sub goals.} \textbf{Total:} 325 \textbf{Proved:} 325

\paragraph{Replay time.} 7 executions timed, removed maximum and minimum. Statistics bellow are from the leftover 5 tests.

\begin{table}
\begin{tabular}{|l|l|l|}
\hline ~\textbf{Minimum}~   & ~\textbf{Maximum}~ & ~\textbf{Average}~~ \\
\hline ~0m29.238s & ~0m30.095s & ~0m29.565s \\ \hline
\end{tabular}
\end{table}

\paragraph{Statistics per prover.} ~

\begin{table}
\begin{tabular}{|l|l|l|l|l|}
\hline
\textbf{Prover}   & \textbf{Number of Proofs} & ~\textbf{Minimum (s)} & \textbf{Maximum (s)} & \textbf{Average (s)}  \\
\hline
Eprover 2.5 &  104 & 0.03  & 4.27 &  0.25 \\
\hline
Alt-Ergo 2.4.0 ~~& 230 & 0.01 &  2.00 &  0.13 \\
\hline
Z3 4.8.7 & 90 &  0.01  & 4.10 &  0.21  \\
\hline
CVC4 1.6  & 97 & 0.07 & 3.56 &  0.25  \\
\hline
\end{tabular}
\end{table}

\end{document}